\documentclass[preprint,12pt,3p]{elsarticle}

\usepackage{amssymb}
\usepackage{amsmath}

\usepackage[framemethod=TikZ]{mdframed}
\usetikzlibrary{shapes.symbols} 
\usetikzlibrary{arrows.meta}
\usetikzlibrary{calc}
\usetikzlibrary{shapes.misc}
\usepackage{xspace}
\usepackage{hyperref}
\usepackage[table]{xcolor}

\newtheorem{definition}{Definition}[section]

\journal{Simulation Modelling Practice and Theory}

\begin{document}

\begin{frontmatter}



\title{Novel Concepts for Agent-Based Population Modelling and Simulation: Updates from GEPOC ABM}

\author[tu_inf]{Martin Bicher} 
\author[tu_cstat,dwh]{Maximilian Viehauser} 
\author[tu_cstat,dwh]{Daniele Giannandrea} 
\author[dwh]{Hannah Kastinger} 
\author[tu_inf,dwh]{Dominik Brunmeir} 
\author[tu_cstat,tu_inf,dwh]{Niki Popper} 

\affiliation[tu_inf]{organization={TU Wien, Institute of Information Systems Engineering, Research Unit Data Science},
            addressline={Favoritenstraße 9-11}, 
            city={Vienna},
            postcode={1040}, 
            state={Vienna},
            country={Austria}}
            
\affiliation[tu_cstat]{organization={TU Wien, Institute of Statistics and Mathematical Methods in Economics, Research Unit Computational Statistics},
addressline={Wiedner Hauptstraße 8-10}, 
city={Vienna},
postcode={1040}, 
state={Vienna},
country={Austria}}

\affiliation[dwh]{organization={dwh GmbH},
addressline={Neustiftgasse 57-59}, 
city={Vienna},
postcode={1070}, 
state={Vienna},
country={Austria}}

\begin{abstract}
In recent years, dynamic agent-based population models, which model every inhabitant of a country as a statistically representative agent, have been gaining in popularity for decision support. This is mainly due to their high degree of flexibility with respect to their area of application. GEPOC ABM is one of these models. Developed in 2015, it is now a well-established decision support tool and has been successfully applied for a wide range of population-level research questions ranging from health-care to logistics. At least in part, this success is attributable to continuous improvement and development of new methods. While some of these are very application- or implementation-specific, others can be well transferred to other population models. The focus of the present work lies on the presentation of three selected transferable innovations. We illustrate an innovative time-update concept for the individual agents, a co-simulation-inspired simulation strategy, and a strategy for accurate model parametrisation. We describe these methods in a reproducible manner, explain their advantages and provide ideas on how they can be transferred to other population models.
\end{abstract}


\begin{highlights}
\item With small changes in model-, simulation- and parametrisation concepts, the well-established agent-based population model GEPOC ABM became much more valid, performant, flexible, and intuitive in the last few years, which enabled various new interesting applications.
\item The individual time-update strategy is now centred around the agent's annual birthday, which came with many advantages compared to the old Jan 1\textsuperscript{st}-centred approach. 
\item Simulation of agent-interactions was inspired by co-simulation, which opens a natural path towards parallelisation.
\item With new tailored disaggregation and probability computation methods, the model can now be parametrised using solely open data sources, with great quantitative validity.
\end{highlights}

\begin{keyword}
modelling and simulation \sep population model \sep agent-based modelling \sep demography



\end{keyword}

\end{frontmatter}




\section{Introduction}
With the continuous increase in computing power, agent-based (AB) population models are becoming increasingly popular as decision support tools. In these models, every agent statistically represents a real person and can be assigned arbitrary attributes and behaviour patterns and can be tracked dynamically over any period of time.

These features give AB population models an almost limitless range of potential applications. Likely, the most prominent field of application is epidemiology. Due to the microscopic nature of AB epidemic models, they allow consideration of details of the real system which cannot be depicted by classic compartmental models \cite{miksch_why_2019}. As a result, they have been heavily used for counselling in the course of the COVID-19 pandemic \cite{rosenstrom_covsim_2024,muller_explicit_2023}. Other applications include the economy \cite{poledna2023}, opinion dynamics \cite{schweitzer2020agent}, and ecology \cite{zhang2020}. Finally, one must not forget potential non-dynamic applications. A snapshot of the agent-population in a population model can be regarded as a synthetic census, which can be used for arbitrary statistical analysis -- a strategy which comes with various benefits~\cite{jarmin_expanding_2014} and drawbacks~\cite{ruggles_shortcomings_2025} compared to actually collected micro-/aggregated- census data.

In short, an agent-based population model can be a highly flexible and valuable tool for decision support. However, the significance of the model results is strongly linked to how accurately the real population and its dynamics are represented in the model. This requires highly accurate modelling and high resolution parametrisation. Moreover, the sheer size of the agent population requires a highly performant implementation to get model results in a reasonable computation time. As a result, an AB population model is often part of a large and well maintained framework, e.g. DYNASIM \cite{favreault_dynamic_2015}, developed for analysis of population wealth, or COVSIM \cite{rosenstrom_covsim_2024}, developed for epidemiological applications. For other examples, we refer to \cite{spielauer_dynamic_2007}.

In 2015, a team at TU Wien and dwh GmbH started the development of their own Generic Population Concept (GEPOC), a collection of models and methods for population modelling and simulation \cite{bicher_definition_2015}, which serves as a sustainable model-base for answering population-level research questions. The most successful and most frequently applied of the models is GEPOC ABM, a generic agent-based population model \cite{bicher_gepoc_2018}. The model is equipped with a default parametrisation for Austria, but could be applied to any country, given that the required data is available.

In the past, GEPOC ABM has been applied to a wide range of applications. These include analysis of vaccination rates, re-hospitalisation of psychiatric patients, number, and severity and diagnosis of stroke incidences (we refer to \cite{bicher_gepoc_2018} for details). Between 2020 and 2023, GEPOC ABM was heavily used for decision support within the COVID-19 crisis. Including previous work on influenza in GEPOC ABM resulted in a highly flexible epidemics model, which became core element of a Covid-19 model-family \cite{bicher_ideas_2025}. The family was used to compute various what-if scenarios and forecasts related to containment policies \cite{bicher_evaluation_2021}, vaccination strategies \cite{bicher_iterative_2022}, detection rate \cite{rippinger_evaluation_2020}, and reporting analysis \cite{popper_synthetic_2020}. Due to the great flexibility of the model, it became one of the key elements in Austria's pandemic management. It was one of the three models harmonised within the Austrian Forecasting Consortium \cite{bicher_supporting_2020} and was one of the approaches in the ECDC European Scenario Hub. At the moment, the model is applied for analysis and economic comparison of 15 different vaccination programs, decision support for climate-change caused disasters such as flooding events \cite{schramm_earth_2025}, analysis of the human papilloma virus (HPV) disease dynamics, and pandemic preparedness \cite{vorstandlechner_impact_2024}.

With ever-changing data and requirements, a tool like GEPOC ABM needs to be regularly revised, revalidated and refactored. This process goes back all the way to the implementation and data-processing to the core of the tool itself: the conceptual model. We found that the latter must satisfy several methodological challenges.

\begin{figure}
     \centering
     \begin{tikzpicture}[
every edge/.append style={nodes={anchor=center, draw,fill=white,font = \footnotesize, inner sep=0.04cm}},
cross/.style={cross out, draw=black, minimum size=9pt, inner sep=0pt, outer sep=0pt,line width=2pt,color=red},
rect_old/.style={rectangle, draw=black, minimum height=7pt, minimum width=0pt, inner sep=0pt, outer sep=0pt},
rect/.style={circle, draw=blue, line width=1pt, minimum width = 5pt,inner sep=0pt, outer sep=0pt},
empty/.style={inner sep=0pt, outer sep=0pt},
circ/.style={draw,circle,line width=2pt, minimum width = 7pt,color=blue}
]
\definecolor{white}{rgb}{1,1,1}
\node[rotate=50,align=left] () at (0.3,1.6) {1929-01-01};
\node[rotate=50,align=left] () at (2.3,1.6) {1930-01-01};
\node[] () at (4.3,1.5) {\dots};
\node[rotate=50,align=left] () at (5.3,1.6) {2000-01-01};
\node[rotate=50,align=left] () at (7.3,1.6) {2001-01-01};
\node[rotate=50,align=left] () at (9.3,1.6) {2002-01-01};
\node[color=blue,rotate=50,align=left] at (1,1.3) {birth};
\node[color=red] at (6.3,-1.9) {event};
\node[rotate=50,color=blue,align=left] at (3.5,1.7) {1\textsuperscript{st} birthday};
\node[rotate=50,color=blue,align=left] at (6.5,1.7) {70\textsuperscript{th} birthday};
\node[rotate=50,color=blue,align=left] at (8.5,1.7) {71\textsuperscript{st} birthday};

\draw[dashed] (0,0.7) to (0,-1.8);
\draw[dashed] (2,0.7) to (2,-1.8);
\draw[dashed] (5,0.7) to (5,-1.8);
\draw[dashed] (7,0.7) to (7,-1.8);
\draw[dashed] (9,0.7) to (9,-1.8);
\draw (1,0.3) node[circ] (A1) {};
\draw (3,0.3) node[rect] (B1) {};
\draw (3.5,0.3) node[empty] (C1) {};
\draw (4.5,0.3) node[empty] (D1) {};
\draw (6,0.3) node[rect] (E1) {};
\draw (8,0.3) node[rect] (F1) {};
\draw (9.3,0.3) node[empty] (G1) {};
\draw (8.3,0.3) node[cross] () {};
\node[] at (12,0.3) {$\Rightarrow$ census $2001$, age $71$};
\path[-Latex,color=blue] (A1) edge (B1);
\path[-,color=blue] (B1) edge (C1);
\draw[-,dotted,color=blue] (C1) to (D1);
\path[-Latex,color=blue] (D1) edge (E1);
\path[-Latex,color=blue] (E1) edge (F1);
\path[-,color=blue] (F1) edge (G1.center);

\draw (1,-0.5) node[circ] (A2) {};
\draw (3,-0.5) node[rect] (B2) {};
\draw (3.5,-0.5) node[empty] (C2) {};
\draw (4.5,-0.5) node[empty] (D2) {};
\draw (6,-0.5) node[rect] (E2) {};
\draw (8,-0.5) node[rect] (F2) {};
\draw (9.3,-0.5) node[empty] (G2) {};
\draw (7.4,-0.5) node[cross] () {};
\node[] at (12,-0.5) {$\Rightarrow$ census $2001$, age $70$};
\path[-Latex,color=blue] (A2) edge (B2);
\path[-,color=blue] (B2) edge (C2);
\draw[-,dotted,color=blue] (C2) to (D2);
\path[-Latex,color=blue] (D2) edge (E2);
\path[-Latex,color=blue] (E2) edge (F2);
\path[-,color=blue] (F2) edge (G2);

\draw (1,-1.3) node[circ] (A3) {};
\draw (3,-1.3) node[rect] (B3) {};
\draw (3.5,-1.3) node[empty] (C3) {};
\draw (4.5,-1.3) node[empty] (D3) {};
\draw (6,-1.3) node[rect] (E3) {};
\draw (8,-1.3) node[rect] (F3) {};
\draw (9.3,-1.3) node[empty] (G3) {};
\draw (6.5,-1.3) node[cross] () {};
\node[] at (12,-1.3) {$\Rightarrow$ census $2000$, age $70$};
\path[-Latex,color=blue] (A3) edge (B3);
\path[-,color=blue] (B3) edge (C3);
\draw[-,dotted,color=blue] (C3) to (D3);
\path[-Latex,color=blue] (D3) edge (E3);
\path[-Latex,color=blue] (E3) edge (F3);
\path[-,color=blue] (F3) edge (G3);
\end{tikzpicture}
     \caption{Visualisation of the \textit{birthday-problem}. Depending on when in the course of the year an event such as birth, migration or death occurs, it will be counted towards different entries in the census data for that year.}
     \label{fig:birthday_problem}
 \end{figure}
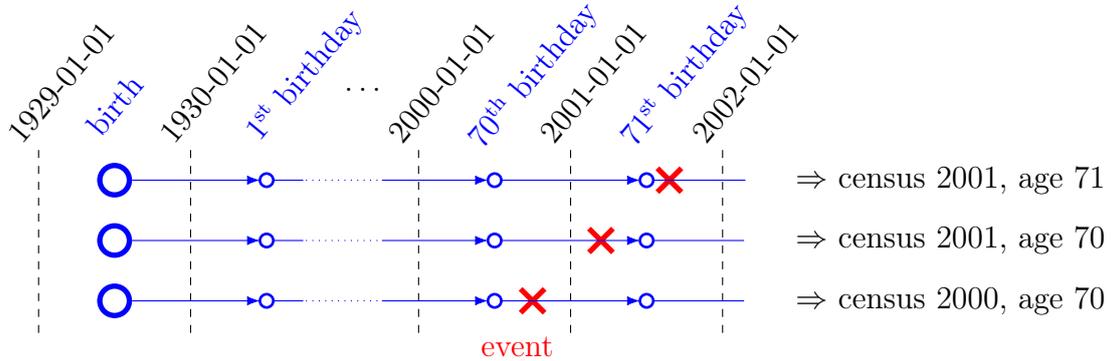

\begin{itemize}
\item \textbf{Flexibility.} A tool such as GEPOC ABM must be flexible with respect to potential applications. Therefore, the conceptual population model must be lightweight, intuitive, and easy to extend. In particular, event-driven updates (continuous-time agent-based modelling) must be supported for applications in which the modeller deals with non-trivial waiting times, e.g. state transitions in disease modelling.
\item \textbf{Performance.} At the same time, the model must also be quick to execute, to avoid bottlenecks for decision support. This problem is often only discussed in terms of implementation, however, it affects model conceptualisation at least in the same magnitude. For example, using an event-based update strategy requires keeping some sort of event-queue sorted all the time throughout the simulation. Independent of the algorithm, the necessary computational efforts will be super-linear with respect to the number of events and consequently also the number of agents -- typically millions in AB population models. Moreover, since any event might change any other future event, options for parallelisation are limited as well.
\item \textbf{Accuracy.} The model must come with a maintainable parametrisation, which is quantitatively validated against reference data. At first glance, census data seems to be a straightforward source for parametrisation of agent behaviour, however, the devil lies in the details as soon as a certain level of accuracy and/or a high level of resolution is required. One good example is the conceptual mismatch between the dynamic age of agents in an AB model and the yearly updated aggregated census data. In a nutshell, the problem is depicted in Figure \ref{fig:birthday_problem}: In case a person experiences a demographic event such as giving birth to a child, dying, or migration, the event will be recorded in the census data. However, it depends on when in the course of the year the event took place, for which age(group) the event will be recorded, that is, if it took place before or after the person's birthday. It is clear that this observation method causes trouble for accurately computing parameters on the individual level. \footnote{Note that this problem might not be present in other cultures, in which ageing works differently. The best example is likely the traditional Korean age, where all people simultaneously age by one at new-year.}
\end{itemize}

Since 2018, when the GEPOC ABM was most recently published~\cite{bicher_gepoc_2018}, several advances have been made, which made the tool more versatile and precise. Some of them are very specific to our implementation and the available data (e.g. the switch from Python3 to ABT/Java in 2019 \cite{brunmeir_four_2023}). 

In this work, however, we want to present and discuss the most relevant and fundamental advancement: a novel strategy for how the model is updated with time. The improvement consists of three parts: an improved update concept for the top-level simulation, a new conceptual model for the individual agents, and an improved parametrisation strategy. The advancement addresses all mentioned modelling challenges at the same time and made our GEPOC ABM not only more flexible, performant and accurate, but also more intuitive and stable. As proof, we will show how the model is currently parametrised with open census data and demonstrate its quantitative validity. We argue that the ideas can be well generalised to any large-scale AB population model and give modellers a new perspective for model conceptualisation.

For particularly interested readers, we refer to \cite{bicher2025gepoc} for a full model and to \cite{bicher_gepoc_2025} for a parameter documentation, which also includes all minor and more specific model updates compared to the last published version in \cite{bicher_gepoc_2018}.

\section{Methods}
\subsection{Modelling and Conceptual Model}
The conceptual model is best described using two layers: the simulation-layer, where agent interaction is modelled, and the agent-layer, where the agent's dynamics are modelled.

Note that for the presentation of the conceptual model, we will focus on clear communication rather than formal correctness and completeness. For the latter, we refer to \cite{bicher2025gepoc}.

\subsubsection{Simulation-layer}
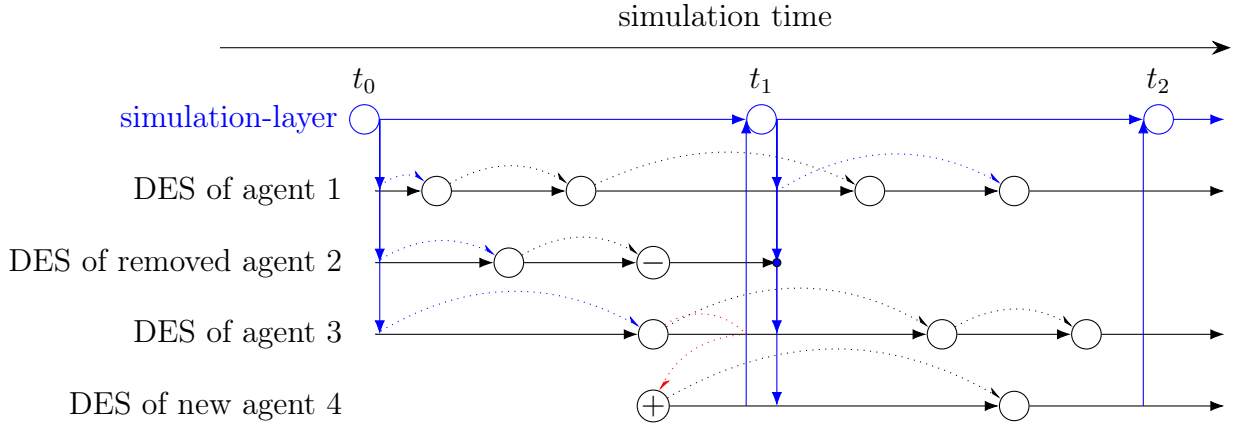
\begin{figure}
    \centering
    \begin{tikzpicture}[scale=0.95,
arrow/.style={-{Latex[scale=1.1]}},
arrow2/.style={dotted, -{Stealth[harpoon]}}
]
\draw (-2,0) edge[-{Stealth[scale=1.5]}] node[pos=0.5,label={above:simulation time}] {} (12,0);
\node[circle,draw=blue,label={[blue]left:simulation-layer},label={above:$t_0$}] (t0) at (0,-1) {};
\node[circle,draw=blue,label={above:$t_1$}] (t1) at (5.5,-1) {};
\node[circle,draw=blue,label={above:$t_2$}] (t2) at (11,-1) {};

\draw (t0.east) edge[arrow,color=blue] (t1.west);
\draw (t1.east) edge[arrow,color=blue] (t2.west);
\draw (t2.east) edge[arrow,color=blue] ($(t2.east)+(0.7,0)$);

\node[label={left:DES of agent $1$}] (al) at (0,-2) {};
\node[circle,draw] (a1) at (1,-2) {};
\node[circle,draw] (a2) at (3,-2) {};
\node[circle,draw] (a3) at (7,-2) {};
\node[circle,draw] (a4) at (9,-2) {};
\draw (al.east) edge[arrow] (a1);
\draw (a1) edge[arrow] (a2);
\draw (a2) edge[arrow] (a3);
\draw (a3) edge[arrow] (a4);
\draw (a4) edge[arrow] ($(t2.east)+(0.7,-1)$);

\node[label={left:DES of removed agent $2$}] (bl) at (0,-3) {};
\node[circle,draw] (b1) at (2,-3) {};
\node[circle,draw,inner sep = 0, minimum width = 9pt] (b2) at (4,-3) {$\large{-}$};
\draw (bl) edge[arrow] (b1);
\draw (b1) edge[arrow] (b2);
\draw (b2) edge[arrow] ($(t1.east)+(0,-2)$);

\node[label={left:DES of agent $3$}] (cl) at (0,-4) {};
\node[circle,draw] (c1) at (4,-4) {};
\node[circle,draw] (c2) at (8,-4) {};
\node[circle,draw] (c3) at (10,-4) {};
\draw (cl.east) edge[arrow] (c1);
\draw (c1) edge[arrow] (c2);
\draw (c2) edge[arrow] (c3);
\draw (c3) edge[arrow] ($(t2.east)+(0.7,-3)$);

\node[label={left:DES of new agent $4$}] (dl) at (0,-5) {};
\node[circle,draw,inner sep = 0, minimum width = 9pt] (d1) at (4,-5) {$+$};
\node[circle,draw] (d2) at (9,-5) {};
\draw (d1) edge[arrow] (d2);
\draw (d2) edge[arrow] ($(t2.east)+(0.7,-4)$);

\draw[color=blue] (t0.east) edge[arrow] ($(t0.east)+(0,-1)$) edge[arrow] ($(t0.east)+(0,-2)$) edge[arrow] ($(t0.east)+(0,-3)$);

\draw[color=blue] (t1.east) edge[arrow] ($(t1.east)+(0,-1)$) edge[arrow] ($(t1.east)+(0,-2)$) edge[arrow] ($(t1.east)+(0,-3)$) edge[arrow] ($(t1.east)+(0,-4)$); 


\draw[color=blue] ($(t1.west)+(0,-4)$) edge[arrow] (t1.west);

\draw[color=blue] ($(t2.west)+(0,-4)$) edge[arrow] (t2.west);


\draw[color=blue] (al.east) edge[arrow2,bend left = 40] (a1);
\draw[color=black] (a1) edge[arrow2,bend left = 30] (a2);
\draw[color=blue] ($(t1.east)+(0,-1)$) edge[arrow2,bend left = 30] (a4);
\draw[color=black] (a2) edge[arrow2,bend left = 25] (a3);

\draw[color=blue] (bl.east) edge[arrow2,bend left = 30] (b1);
\draw[color=black] (b1) edge[arrow2,bend left = 30] (b2);

\draw[color=blue] (cl.east) edge[arrow2,bend left = 30] (c1);
\draw[color=red] (c1) edge[dotted,bend left = 40] ($(t1.west)+(0,-3)$);
\draw[color=black] (c1) edge[arrow2,bend left = 30] (c2);
\draw[color=black] (c2) edge[arrow2,bend left = 30] (c3);

\draw[color=red]  ($(t1.west)+(0,-3)$) edge[arrow2, bend right = 30] (d1);
\draw[color=black] (d1) edge[arrow2, bend left = 30] (d2);

\node[fill=blue,draw, circle,inner sep = 0, minimum width = 3pt]  at ($(t1.east)+(0,-2)$) {};
\end{tikzpicture}
    \caption{Overall update scheme of GEPOC ABM. The time advancement is sketched with solid horizontal arrows, event-scheduling with dotted harpoons. The simulation-layer acts as runtime-infrastructure for the discrete-event simulations (DESs) of the individual agents. In every macro step, the simulation-layer (blue) first observes the individual states of the DESs (upwards arrows) and potentially interferes with them (downwards arrows) which includes adding or removing events (blue dotted arrows) or entire agents. The red dotted arrow displays an interaction between agents via the simulation-layer -- in this case, a ``birth''-event.}
    \label{fig:co_simulation_concept}
\end{figure}

To make an AB population model with millions of agents computable with event-based update, we specified a very specific update strategy, which can best be described as a form of co-simulation~\cite{hafner_overview_2021}. In this approach, every agent itself follows its own individual discrete-event simulation (DES), which are harmonised by a synchronisation reference, henceforth called the \textit{simulation-layer}. As in traditional parallel co-simulation, the simulation-layer uses macro steps on which it synchronizes the individual DESs (see Figure \ref{fig:co_simulation_concept}). At every step, the simulation-layer first observes the individual DESs and tracks whether an agent (a) experienced an event which renders it \textit{removed} in the course of the last macro step, (b) scheduled an event which should lead to the creation of a new agent, or (c) scheduled an event which addresses another existing agent. In case of (a) the simulation-layer would remove the corresponding agents from the harmonisation loop, in case of (b) it would initialise a new agent and the corresponding DES and would trigger its initialisation event. In case of (c), it would insert the event into the event-queue of the other agent \footnote{Note that removal and addition of new simulation entities is usually not an element of traditional co-simulation}. While this strategy leaves interaction between agents restricted to the macro-steps, the individual agents themselves can be updated time continuously. However, with a proper choice of the macro-step-size this limitation can be overcome, depending on the application. Reasonable values range from ``seconds'' for applications in mobility, ``days'' for epidemiological applications, to ``years'' for applications in environmental sciences. As compensation for the limited interaction possibilities, one obtains a model that combines discrete and event-based updates, evaluates quickly, and enables direct application of parallelisation at the agent level. We refer to the Discussion section for details. 

Finally, we want to emphasize that also the simulation-layer in GEPOC ABM is implemented as its own DES, where macro-steps are simply events scheduled iteratively. This leaves the space open for arbitrary non-equidistant time steps such as ``months'' or ``years'', and enables a more modular structure when adding additional top-level events, e.g. for observation or runtime plotting.

\subsubsection{Agent-layer}
The agents themselves are modelled as traditional discrete event simulations with extended perception of their environment, i.e. the other agents. For modelling demographics, the internationally acknowledged definition of the term \textit{probability of death}, as it is used by various national census institutes, was the key inspiration. 
\begin{definition}[Probability of Death]\label{def:deathprob}
``The probability of death at some age x refers to the probability of a person living until the age of x to die during that year of age.'' (Online glossary of Statistics Finland~ \cite{statistics_finland_probability_nodate})
\end{definition}
The most interesting feature of this definition is how it is centred around an individual’s year of life, and not around the calendar year. Since the definition is international standard, there are various methods for its interplay with the population census, which can be used for parametrisation (see below). Therefore, we defined an event-based update-scheme built on the same principle.

Let $a^i$ stand for model-agent $i$ and let $a^i(t)$ stand for the agent's state vector at time $t$. Hereby, $\left(a^i(t)\right)_1:=bd$ refers to the static birthdate of the agent, $\left(a^i(t)\right)_2:=age$ refers to its age in years, and $\left(a^i(t)\right)_j,j=3,\dots,n$ refer to other static or dynamic features of the agent such as sex, region/place of residence, annual income, or co-morbidities. In order to present the conceptual update scheme, we do not need to specify them any further.

\begin{figure}
    \centering
    \begin{tikzpicture}[
eventtype/.style = {circle, draw, minimum width = 2 cm,text=black,draw=black, align = center},
interface/.style = {rectangle, draw, minimum width = 2 cm,text=black, align = center},
scheduleplain/.style = {-{Stealth[scale=1.5]}},
nodeif/.style = {pos=0.66,anchor=center, circle, minimum width=0.5cm},
nodetime/.style = {pos=0.05,anchor=center},
nodearg/.style = {pos=0.4,anchor=center, rectangle, minimum width=0.5 cm,fill=white,draw},
]

\node[eventtype,label={[align=left]below:$a^i(t)\leftarrow (bd,0,\dots)$}] (init) at (1.6,0) {\textit{Init.}};

\node[eventtype,label={[align=left]below:$a^i(t)_2++$}] (bd) at (5,0) {\textit{Birthday}};

\draw (init) edge[scheduleplain] node[nodetime,pos=0.5,label ={above:$\Delta_{t,bd}$}] {} (bd);

\draw (bd) edge[scheduleplain,out = 50, in = 130, looseness = 3] node[nodetime,pos=0.3,label ={above:$\Delta_{t,bd}$}] {} (bd);

\node[eventtype,label ={below:$a^i(t)\leftarrow \dots$}] (ev) at (10,0) {\textit{Event}};

\node[eventtype,dotted] (ev2) at (13.5,3) {\dots};

\draw[scheduleplain,dotted,bend left = 20] (ev) to (ev2);

\draw (bd) edge[scheduleplain] node[nodetime,pos=0.3,label ={below:$\Delta_{t,bd}\cdot U_{0,1}$}] {} node[nodeif, pos=0.6,label={above:$p(t,a^i(t))$}] {\large{$\int$}} (ev);

\draw (init) edge[scheduleplain,out=70,in=120] node[nodetime,pos=0.35,label ={ above:$\Delta_{t,bd}\cdot U_{0,1}$}] {} node[nodeif, pos=0.6,label={above:$\frac{\Delta_{t,bd}}{\Delta_{t,bd}+\Delta_{bd^{-},t}}\cdot p(t-\Delta_{bd^-,t},\tilde{a}^i)$},rotate=350] {} node[pos=0.7] {\large{$\int$}} (ev);

\node[eventtype,label ={below:$a^i(t)\leftarrow null$}] (rem) at (13.5,0) {\textit{Remove}};

\draw[scheduleplain] (ev) to node[nodeif, pos=0.5,label={[align=center]above:{event is\\terminal?}}]{\large{$\int$}} (rem);

\draw[scheduleplain,dashed,bend left = 40] (rem) to (ev);
\draw[scheduleplain,dashed,bend left = 40] (rem) to (bd);
\end{tikzpicture}
    \caption{Event-graph representation of the birthday-centred update-scheme for agents. A \textit{Birthday} event regularly schedules itself every year. On its occurrence, demographic events (e.g. \textit{Event}) taking place in the course of the agent's next life-year will be planned. They will be scheduled with a certain probability at a random point within the next life-year.}
    \label{fig:update_agent}
\end{figure}
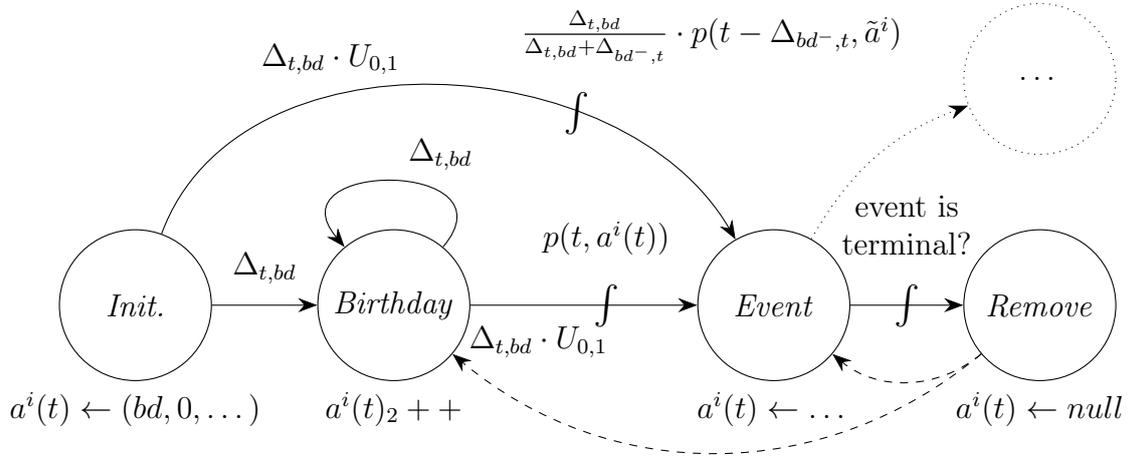

With this definition, the update of the agent follows the Event-Graph depicted in Figure \ref{fig:update_agent} (see \cite{schruben_simulation_1983} for a specification of the Event Graph formalism). The \textit{Birthday} event is the core of the update scheme. It iteratively schedules itself every year on the re-occurrence of the agent's birthday and increments the agent's age (i.e. $a^i(t)_2$) by one. Hereby, in the graph, $\Delta_{t,bd}$ always refers to the time between the current time $t$ and the corresponding next (re-)occurrence of the agent's birthday. It should be noted that this computation is not necessarily trivial, since ``1-year'' is not a proper unit, primarily due to leap-years.

Upon occurrence of the \textit{Birthday} event, time $t$ and state $a(t)$ dependent probabilities decide if a demographic event such as death, birth, migration, etc. should occur within the upcoming life-year. In Figure \ref{fig:update_agent}, this is illustrated by an example event \textit{Event}, which occurs with probability $p(t,a^i(t))$. In case the event is to be scheduled, it will be planned for a random point in time between the current and next \textit{Birthday} event. That means, a time-delay is computed by multiplication of $\Delta_{t,bd}$ with a random number $U_{0,1}\sim U(0,1)$. We also want to emphasise that the \textit{Birthday} event also increases the agents' age by one. That means, the agent's age in years does not have to be (re-)computed from the difference between the current year and the agent's day of birth -- as it would be standard in other implementations -- but comes ``for-free'' as a dynamic agent state. 

When a demographic event occurs, it can change the state $a(t)$; e.g., an internal migration event changes the place of residence. The event itself can also trigger further events; e.g., a birth event initializes a new agent. In case an event is terminal, e.g. a death-event, the \textit{Remove} event will be scheduled, which (a) renders the agent as inactive, which is indicated in the Event Graph by setting the state to \textit{null}, and (b) cancels all further planned events, indicated by the dashed cancelling edges.

In the model, a new agent is created when the initial agent population is initialised, when an agent ``gives birth'' to a new agent at runtime, or when an agent enters the system via immigration processes. In all three cases, an \textit{Init.} event sets up the initial state of the agent and schedules the first birthday event with delay $\Delta_{t,bd}$. Note that this time-span will be one year if the agent was born at runtime, otherwise it will be shorter. In both cases, there is a chance that a demographic event will take place before the agent ``celebrates'' its first birthday at simulation runtime. The corresponding likelihood is, however, scaled down by $\Delta_{t,bd}/(\Delta_{t,bd}+\Delta_{bd^-,t})$. Hereby, $\Delta_{bd^-,t}$ refers to the time between the agent's last birthday and $t$, i.e. $\Delta_{t,bd}+\Delta_{bd^-,t}$ is precisely the length of the agent's current life-year. There is, however, the problem that the corresponding probability function $p$ must also be evaluated on the day of the agent's last birthday $t-\Delta_{bd^-,t}$. This requires finding an extrapolation $\tilde{a}^i$ for the unknown $a^i(t-\Delta_{bd^-,t})$, e.g. by $\tilde{a}^i\approx a^i(t)$.

In the most fundamental version, GEPOC-ABM uses three demographic agent events: \textit{Death} and \textit{Emigration} represent \textit{terminal} events that remove the agent from the model. \textit{Birth} leads to the creation of a new agent via the simulation-layer. Note that immigration is performed externally by a different agent (Interface Agent), which we will not discuss any further in this work, but refer to \cite{bicher2025gepoc}. In a more advanced version, also an \textit{Internal Migration} event is possible, which models the internal migration of an agent changing its place of residence. Incorporating internal migration improves model accuracy but increases complexity and computational cost. Moreover, since there are versions of GEPOC ABM which do not include any spatial dimension, modelling of internal migration is not always necessary (see below, Section \ref{sec:impl}).

It should be noted that GEPOC ABM differentiates between male and female agents from the biological perspective, since it is primarily used to differentiate reproductive capacity, that is, \textit{Birth} events are only scheduled for female agents. This choice is also reasonable from the parametrisation perspective since it is also the most common concept in census statistics. However, if a specific application requires the use of additional social genders, the agents must be extended by the corresponding attributes.

\subsection{Model Implementation}
\label{sec:impl}
Since 2019, the model has been implemented in Java using the Agent-Based Template (ABT, see \cite{brunmeir_four_2023}). The framework offers sufficient degrees of freedom to support any application/extension of GEPOC ABM, while at the same time providing efficient default methods for things like random number generation, time/date conversions, IO-routines, serialisation and deserialisation of agents, and event-queueing.

\subsection{Parametrisation and Validation Strategy}
GEPOC ABM is only as valid as the used parameters, which are the size and distribution of the initial population, the number and distribution of immigrants entering the simulation at runtime, and the probabilities for demographic events mentioned earlier.

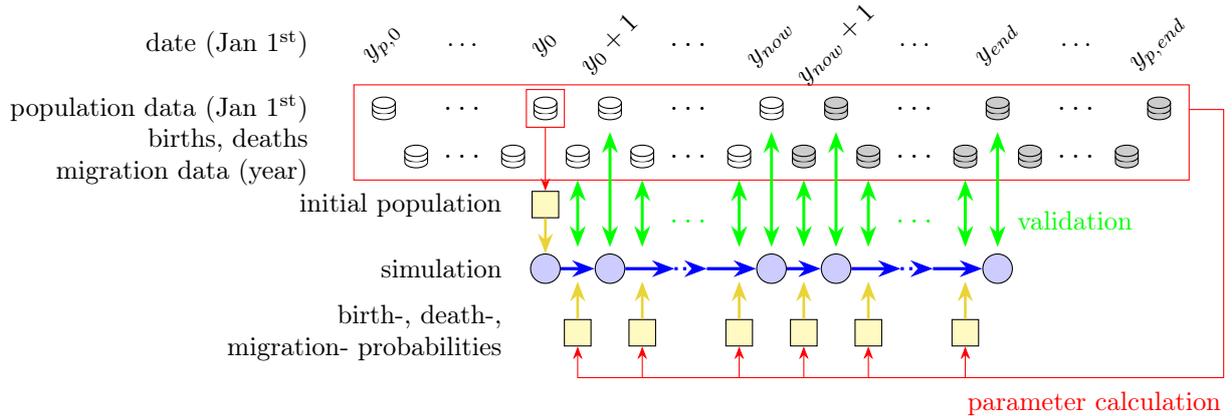
\begin{figure}
    \centering
    \begin{tikzpicture}[scale=0.85, database/.style={
        inner xsep = 0.0,
        outer xsep = 0.0,
        path picture={
            \fill[fill=#1] (-0.5,-0.25) rectangle (0.5,0.25);
            \draw[fill=#1] (0.0, 0.25) circle [x radius=0.5,y radius=0.25];
            \draw (-0.5, 0.0) arc [start angle=180, end angle=360,x radius=0.5, y radius=0.25];
            \draw[fill=#1] (-0.5,-0.25) arc [start angle=180, end angle=360,x radius=0.5, y radius=0.25];
            \draw (-0.5,-0.25) -- (-0.5,0.25);
            \draw (0.5,-0.25) -- (0.5,0.25);
        },
        minimum width=1.5cm,
        minimum height=1.5cm,
        fill=white,
        scale=0.3
    },
    simstate/.style={circle,draw, fill=blue!20,minimum width=0.35cm,
        minimum height=0.35cm},
    simparam/.style={draw,fill=yellow!30,minimum width=0.35cm,
        minimum height=0.35cm},
    paramarrow/.style = {-Stealth,draw=yellow!90!blue,line width = 1},
    simdyn/.style={-Stealth,draw=blue, line width = 1.2},
    labelright/.style={align=right, anchor=east,font=\footnotesize},
    parametercomp/.style={draw=red},
    validation/.style={Stealth-Stealth,draw=green,line width=1.1}
    ]


\node[labelright] at (-1,1) {date (Jan 1\textsuperscript{st})};
\node[rotate=45,font=\footnotesize] at (0,1) {$y_{p,0}$};
\node[font=\footnotesize] at (1.25,1) {$\dots$};
\node[rotate=45,font=\footnotesize] at (2.5,1) {$y_{0}$};
\node[rotate=45,font=\footnotesize] at (3.5,1) {$y_{0}+1$};
\node[] at (4.75,1) {$\dots$};
\node[rotate=45,font=\footnotesize] at (6,1) {$y_{now}$};
\node[rotate=45,font=\footnotesize] at (7,1) {$y_{now}+1$};
\node[font=\footnotesize] at (8.25,1) {$\dots$};
\node[rotate=45,font=\footnotesize] at (9.5,1) {$y_{end}$};
\node[font=\footnotesize] at (10.75,1) {$\dots$};
\node[rotate=45,font=\footnotesize] at (12,1) {$y_{p,end}$};

\node[labelright] at (-1,0) {population data (Jan 1\textsuperscript{st})};
\node[database={white}] (p1) at (0,0) {};
\node[] at (1.25,0) {$\dots$};
\node[database={white}] (p2) at (2.5,0) {};
\node[database={white}] (p3) at (3.5,0) {};
\node[] at (4.75,0) {$\dots$};
\node[database={white}] (p4) at (6,0) {};
\node[database={black!20}] (p5) at (7,0) {};
\node[] at (8.25,0) {$\dots$};
\node[database={black!20}] (p6) at (9.5,0) {};
\node[] at (10.75,0) {$\dots$};
\node[database={black!20}] (p7) at (12,0) {};

\node[labelright] at (-1,-0.75) {births, deaths\\ migration data (year)};
\node[database={white}] (d1) at (0.5,-0.75) {};
\node[] at (1.25,-0.75) {$\dots$};
\node[database={white}] (d2) at (2,-0.75) {};
\node[database={white}] (d3) at (3,-0.75) {};
\node[database={white}] (d4) at (4,-0.75) {};
\node[] at (4.75,-0.75) {$\dots$};
\node[database={white}] (d5) at (5.5,-0.75) {};
\node[database={black!20}] (d6) at (6.5,-0.75) {};
\node[database={black!20}] (d7) at (7.5,-0.75) {};
\node[] at (8.25,-0.75) {$\dots$};
\node[database={black!20}]  (d8) at (9,-0.75) {};
\node[database={black!20}]  (d9) at (10,-0.75) {};
\node[] at (10.75,-0.75) {$\dots$};
\node[database={black!20}] (d10) at (11.5,-0.75) {};

\node[labelright] at (2,-2.5) {simulation};
\node[simstate] (s1) at (2.5,-2.5) {};
\node[simstate] (s2) at (3.5,-2.5) {};
\coordinate (s21) at (4.5,-2.5) {};
\coordinate (s22) at (5,-2.5) {};
\node[simstate] (s3) at (6,-2.5) {};
\node[simstate] (s4) at (7,-2.5) {};
\coordinate (s41) at (8,-2.5) {};
\coordinate (s42) at (8.5,-2.5) {};
\node[simstate] (s5) at (9.5,-2.5) {};
\draw[simdyn] (s1) to (s2);
\draw[simdyn] (s2) to (s21);
\draw[simdyn,dotted] (s21) to (s22);
\draw[simdyn] (s22) to (s3);
\draw[simdyn] (s3) to (s4);
\draw[simdyn] (s4) to (s41);
\draw[simdyn,dotted] (s41) to (s42);
\draw[simdyn] (s42) to (s5);

\node[labelright] at (2,-1.5) {initial population};
\node[simparam] (c1) at (2.5,-1.5) {};
\draw[paramarrow] (c1) to (s1);

\node[simparam] (c2) at (3,-3.5) {};
\node[simparam] (c3) at (4,-3.5) {};
\node[simparam] (c4) at (5.5,-3.5) {};
\node[simparam] (c5) at (6.5,-3.5) {};
\node[simparam] (c6) at (7.5,-3.5) {};
\node[simparam] (c7) at (9,-3.5) {};

\node[labelright] at (2,-3.5) {birth-, death-, \\migration- probabilities};
\draw[paramarrow] (c2) to ++ (0,0.8);
\draw[paramarrow] (c3) to ++ (0,0.8);
\draw[paramarrow] (c4) to ++ (0,0.8);
\draw[paramarrow] (c5) to ++ (0,0.8);
\draw[paramarrow] (c6) to ++ (0,0.8);
\draw[paramarrow] (c7) to ++ (0,0.8);

\node[parametercomp,minimum width=0.5cm,minimum height=0.5cm] (pp2) at (p2) {};
\draw[parametercomp,-Stealth] (pp2) to (c1);
\draw[parametercomp] ($(p1.north west)+(-0.2,0.1)$) rectangle ($(d10.south east)+(0.7,-0.1)$);
\draw[parametercomp,-Stealth] ($(d10.south east)+(0.7,1.0)$) -- ($(d10.south)+(1.5,1.0)$) -- ($(c7)+(4,-0.7)$) -- ($(c7)+(0,-0.7)$) -- (c7);
\draw[parametercomp,-Stealth] ($(c7)+(0,-0.7)$) -- ($(c6)+(0,-0.7)$) -- (c6);
\draw[parametercomp,-Stealth] ($(c6)+(0,-0.7)$) -- ($(c5)+(0,-0.7)$) -- (c5);
\draw[parametercomp,-Stealth] ($(c5)+(0,-0.7)$) -- ($(c4)+(0,-0.7)$) -- (c4);
\draw[parametercomp,-Stealth] ($(c4)+(0,-0.7)$) -- ($(c3)+(0,-0.7)$) -- (c3);
\draw[parametercomp,-Stealth] ($(c3)+(0,-0.7)$) -- ($(c2)+(0,-0.7)$) -- (c2);

\draw[validation] ($(s1.north)+(0.5,0.1)$) -- ($(d3.south)+(0,-0.1)$);
\draw[validation] ($(s2.north)+(0,0.1)$) -- ($(p3.south)+(0,-0.1)$);
\draw[validation] ($(s2.north)+(0.5,0.1)$) -- ($(d4.south)+(0,-0.1)$);
\draw[validation] ($(s3.north)+(-0.5,0.1)$) -- ($(d5.south)+(0,-0.1)$);
\draw[validation] ($(s3.north)+(0,0.1)$) -- ($(p4.south)+(0,-0.1)$);
\draw[validation] ($(s4.north)+(-0.5,0.1)$) -- ($(d6.south)+(0,-0.1)$);
\draw[validation] ($(s4.north)+(0,0.1)$) -- ($(p5.south)+(0,-0.1)$);
\draw[validation] ($(s4.north)+(0.5,0.1)$) -- ($(d7.south)+(0,-0.1)$);
\draw[validation] ($(s5.north)+(-0.5,0.1)$) -- ($(d8.south)+(0,-0.1)$);
\draw[validation] ($(s5.north)+(0.0,0.1)$) -- ($(p6.south)+(0,-0.1)$);
\node[text=green] at (4.75,-1.75) {$\dots$};
\node[text=green] at (8.25,-1.75) {$\dots$};
\node[text=green,font=\footnotesize] at (10.7,-1.75) {validation};
\node[text=red,font=\footnotesize] at (11,-4.6) {parameter calculation};
\end{tikzpicture}
    \caption{General parametrisation and validation scheme of GEPOC ABM. Population, death, birth and migration counts in form of census counts (white) and forecasts (grey) pose the source for the parameter values and the validation reference at the same time.}
    \label{fig:validation}
\end{figure}

To enable the highest amount of flexibility, GEPOC ABM is based on semi-automatically updated parametrisation data, which is computed from openly available sources. For Austria, which is the main area of application of GEPOC ABM, the parametrisation data is computed from 14 different openly available datasets, including historic census and forecasting data for various demographic indicators, as well as geographical data (shapes and raster data) for regional sampling. The most important source of data is the official national statistics bureau Statistics Austria and its regularly updated open data platform \href{https://data.statistik.gv.at/web/}{https://data.statistik.gv.at/web/}. In GEPOC ABM, the data is used as source for parameters and as validation reference at the same time. This is displayed in Figure \ref{fig:validation}. Depending on whether the start date $y_0$ of the simulation is in the past, present, or future,  the simulation is initialised either with the corresponding population counts from the census data or from the official forecast of the national statistics office. In Figure \ref{fig:validation}, the simulation was started in the past, since $y_0<y_{now}$. From this point forward, the simulation is driven by probability parameter values, which are computed from the data ex ante. For validation purposes, the simulation computes a synthetic census by counting living, died, newborn, emigrated and immigrated agents. The comparison between the simulated and the actual data (census and forecast) constitutes the validation strategy.

Since forecasted information is also used as validation reference, the validation can, at least partially, be interpreted as a form of cross-model-validation, in this particular case against the STATSIM model applied by Statistics Austria~\cite{pohl_statsim_2024}. 
In the validation process, besides modelling errors, also parametrisation errors have to be expected, since the parameter computation process involves several challenges:

\subsubsection{Resolution Mismatch}
The most obvious challenge is the harmonisation of two datasets for two different time-spans with different resolutions, for example, merging recent census data with very high spatial and age resolution with low-resolution forecasts. To solve the mismatch, we apply classic statistical data disaggregation onto the highest degree of resolution, using reasonable assumptions for the underlying distribution, usually taken from the temporarily closest available high-resolution data-set.

Let, for example, $P(2025,[10,19])$ refer to the population with age between 10 and 19 at 2025-01-01, and assume that data with high age resolution is available for 2024, i.e. $P(2024,[i,i])$ for $i\in \{10,\dots,19\}$, then we may compute a distribution vector $\vec{p}_i:=\frac{P(2024,[i,i])}{\sum_{j=10}^{19}P(2024,[j,j])}$ and disaggregate
\begin{equation}
    \forall i\in \{10,\dots,19\}:\ \tilde{P}(2025,[i,i]):=P(2025,[10,19])\vec{p}_i.
\end{equation}
In some cases, the disaggregation must result in whole numbers., e.g. for the computation of the initial population size. For those cases, a variant of the Huntington-Hill method \cite{balinski_huntington_1977} is applied. This method is originally applied for fair distribution of parliament seats with respect to election results, however the concept perfectly applies to disaggregation tasks with integer constraints.

Internal migration data turned out to be a special case of this problem. The official census data provides age-dependent counts for internal emigrants and immigrants, and inter-regional flows between the regions without age resolution. The three datasets can be interpreted as marginals of an unknown internal migration census, whereas the sum over one of the three dimensions, origin, destination and age, must result in one of the three datasets. In order to fulfil this disaggregation task, we apply three-dimensional Iterative Proportional Sampling (see, \cite{deming1940least}). Based on an initial guess, this algorithm iteratively scales its internal state-tensor by the three defined marginals. In case the system has a valid solution, the algorithm converges after sufficiently many scaling steps. The resulting data can be used for an internal migration model, which preserves valid origin-destination flows \underline{and} the correct age-distribution for internal immigrants. Note that the resulting origin-destination flows per age group are real numbers and not integers, and must therefore be understood as a distribution rather than a person count.

Finally, we want to emphasise that the resolution-matching tasks for the open-data parametrisation of GEPOC ABM can range in complexity from the mentioned disaggregation of age-classes into single age groups, to the computation of high-resolution yearly deaths or births from scalar demographic indicators like life-expectancy or mean-fertility age.

\subsubsection{Probabilities from Census Data}
As mentioned earlier, the computation of probabilities from classical census data is always problematic. However, with the applied modelling strategy, in which the probabilities are based on Definition \ref{def:deathprob}, we can make use of existing, validated methods for their computation. The most important of these originates in the computation of life-tables and was developed by life-science pioneer William Farr (1807-1883).

To illustrate this, we take the computation of death probabilities as example. Let $P(y,a)$ stand for the population per (turn of the) year $y$, and age $a$, and $D(y,a)$ for the overall number of deaths of $a$-year old individuals in year $y$ (we drop any other dimensions such as sex and region for the sake of readability). 

It is clear that naive computation of the average rate of mortality $m=D(y,a)/P(y,a)$ already provides a reasonable approximation to the probability of death $p_d(y,a)$, however, it does not properly regard the mismatch between census data year and individual life year. Note that roughly half of the died individuals which contributed to $D(y,a)$ were actually $a-1$ years old at the beginning of year $y$. Farr's death rate formula (see \cite{farr_construction_1859}, page 848) takes this problem into account:
\begin{equation}p_d(y,a)\approx 1-\frac{1-\frac{1}{2}\frac{D(y,a)}{P_{avg}(a)}}{1+\frac{1}{2}\frac{D(y,a)}{P_{avg}(a)}}=\frac{D(y,a)}{P_{avg}(a)+\frac{1}{2}D(y,a)}.\end{equation}
The formula assumes a (temporarily) constant rate of mortality, a constant age-cohort size $P_{avg}(a)$, and neglects migration. The formula can be reasoned by the idea that the average rate of mortality $m$ computes the rate of death per lived year of life, but not per individual. Precisely $P_{avg}(a)$ years are lived by persons with age $a$ in the course of the observed year, however, since every died individual will instantaneously be replaced by a new one (e.g. from the next age cohort) to maintain the constant size of the age cohort, those years are lived by $P_{avg}(a)+D(y,a)$ distinct persons.

 By default, all parameters are computed on different regional aggregation levels (country, federal-states, districts, municipalities). First, this concept provides the highest possible level of flexibility with respect to the model application. For example, it is possible to completely neglect any regional resolution if not needed and save valuable computation time. Furthermore, the concept also enables the selection of a suitable regional resolution for every demographic process individually. For example, in Austria, emigration probabilities are regionally highly diverse, and therefore the data with the highest available regional resolution should be used. On the contrary, it turned out to be not only sufficient but also beneficial to consider death probability parameter values on the federal-state resolution, since the parameter computation is more stable due to the larger population and death counts.
 
Currently, the established parameter-files cover the maximum time period between 1996 and 2101. For a full specification of the parametrisation process, we refer to \cite{bicher_gepoc_2025}.

\section{Results}
\subsection{Validation Scenarios}
For the validation, we define 2000-01-01 as the start-date and 2050-01-01 as the end date of the simulation and use GEPOC to dynamically update the population of Austria. That means, the model uses the population census data for 2000-01-01, the death, birth and migration parameters for the years 2000 to 2024 and the corresponding forecasts between 2025 and 2049. We choose yearly macro steps and run the model under full consideration of internal migration (IM) -- using the terminology of \cite{bicher2025gepoc}, we apply the Full-Regional IM Model, which uses age as well as origin and destination resolved internal migration parameters. The probability parameters defined in the previous sections are regarded as year-, age-, sex-, and region-dependent, whereas different resolutions are used for the latter: birth- and death-probabilities are different for every federal-state (9), migration-probabilities are different for every district (roughly 100).

Over the 50 simulated years, the simulation develops its own synthetic census by counting the total number of inhabitants (on each Jan 1\textsuperscript{st}) and the numbers of births, deaths, and emigrants each year. For comparison with the actual data, we observe the synthetic-census for the  federal-states of Austria, and by 20-year age classes. Hence, the output of a simulation run is defined as the four functions
\begin{multline}
    P^{sim}:\{2000,\dots,2050\}\times \{\text{AT-1},\dots,\text{AT-9}\}\times \{m,f\}\times \{[0,19],\dots,80+\}\rightarrow \mathbb{N}:\\
    (y,r,s,a)\mapsto P^{sim}(y,r,s,a)
\end{multline}
for the total population on $y-01-01$ and with $X\in \{B,D,E\}$,
\begin{multline}
    X^{sim}:\{2000,\dots,2049\}\times \{\text{AT-1},\dots,\text{AT-9}\}\times \{m,f\}\times \{[0,19],\dots,80+\}\rightarrow \mathbb{N}:\\
(y,r,s,a)\mapsto X^{sim}(y,r,s,a)
\end{multline}
for total births, deaths and emigrants over the year $y$. Validation of the immigrants is not necessary, since they are directly and deterministically sampled from the census data.

Since they are outputs of a stochastic simulation, all functions $P,B,D,E$ must be regarded as random numbers. To account for this uncertainty, 9 Monte Carlo runs have been found to be fully sufficient, so that the sample mean 
\begin{equation}
    \overline{X}^{sim}(y,r,s,a):=\frac{1}{9}\sum_{i=1}^{9}X_i^{sim}(y,r,s,a),
\end{equation}
for $X\in \{P,B,D,E\}$ and $X_i^{sim}(y,r,s,a)$ being the result of the $i$-th simulation run, approximates the actual unknown mean $\mu(X^{sim})$ sufficiently well, at least for all $(y,r,s,a)$ with $\overline{X}^{sim}(y,r,s,a)\geq 1000$. This was evaluated using a Gaussian error assumption (compare with \cite{bicher_review_2019}) and by investigating the quotient between sample standard-deviation and sample mean.

The calibration reference $X^{data}$, with $X \in \{P,B,D,E\}$, is a composite of different open data sources, which mainly originate from the Austrian national census office Statistics Austria and its Open Data platform \cite{noauthor_statistik_nodate} (we refer to the corresponding section in \cite{bicher_gepoc_2025} for details). As displayed in Figure \ref{fig:validation}, the simulation results are compared with available census counts between 2000 and 2025 and with the forecasting data between 2026 and 2050. Since birth, death and migration forecasts do not include information about age and sex, a higher level of aggregation is chosen.

As validation metric, we define the maximum and minimum relative deviation $e_{max},e_{min}$ for a given time interval $y_1,y_2$, region $r$, sex $s$ and age $a$ as
\begin{equation}
    e_{max}(\overline{X}^{sim},r,s,a):=\max_{y\in\{y_1,\dots,y_2\}}\left(\frac{\overline{X}^{sim}(y,r,s,a)-X^{data}(y,r,s,a)}{\max(1,X^{data}(y,r,s,a))}\right),
\end{equation}
\begin{equation}
    e_{min}(\overline{X}^{sim},r,s,a):=\min_{y\in\{y_1,\dots,y_2\}}\left(\frac{\overline{X}^{sim}(y,r,s,a)-X^{data}(y,r,s,a)}{\max(1,X^{data}(y,r,s,a))}\right).
\end{equation}
This validation metric, which, to be mathematically precise, is not a metric but rather an error interval, was chosen in favour of classic errors like mean-squared error, because it does not smooth over fluctuations and maintains the sign of the difference. As a result, potential errors due to temporary fluctuations and systematic under- or overestimation remain visible. 

Since the interval between the 5 and 95 percent sample quantile \\ $\left(Q^{0.05}(X^{sim}),Q^{0.95}(X^{sim})\right)$ can be regarded as $90\%$ confidence interval for $X^{sim}$, so can
\begin{equation}
\left(e_{max}(Q^{0.05}(X^{sim}),r,s,a),e_{max}(Q^{0.95}(X^{sim}),r,s,a)\right)
\end{equation}
for the maximum deviation, minimum deviation analogously. While we will state this confidence band in the result tables, we will omit drawing confidence bands in the result plots. The latter only display highly aggregated system components, for which the uncertainty is very low and the corresponding confidence band would be almost invisible.

\subsection{Validation Results}
\begin{figure}
    \centering
    \includegraphics[width=1.0\linewidth]{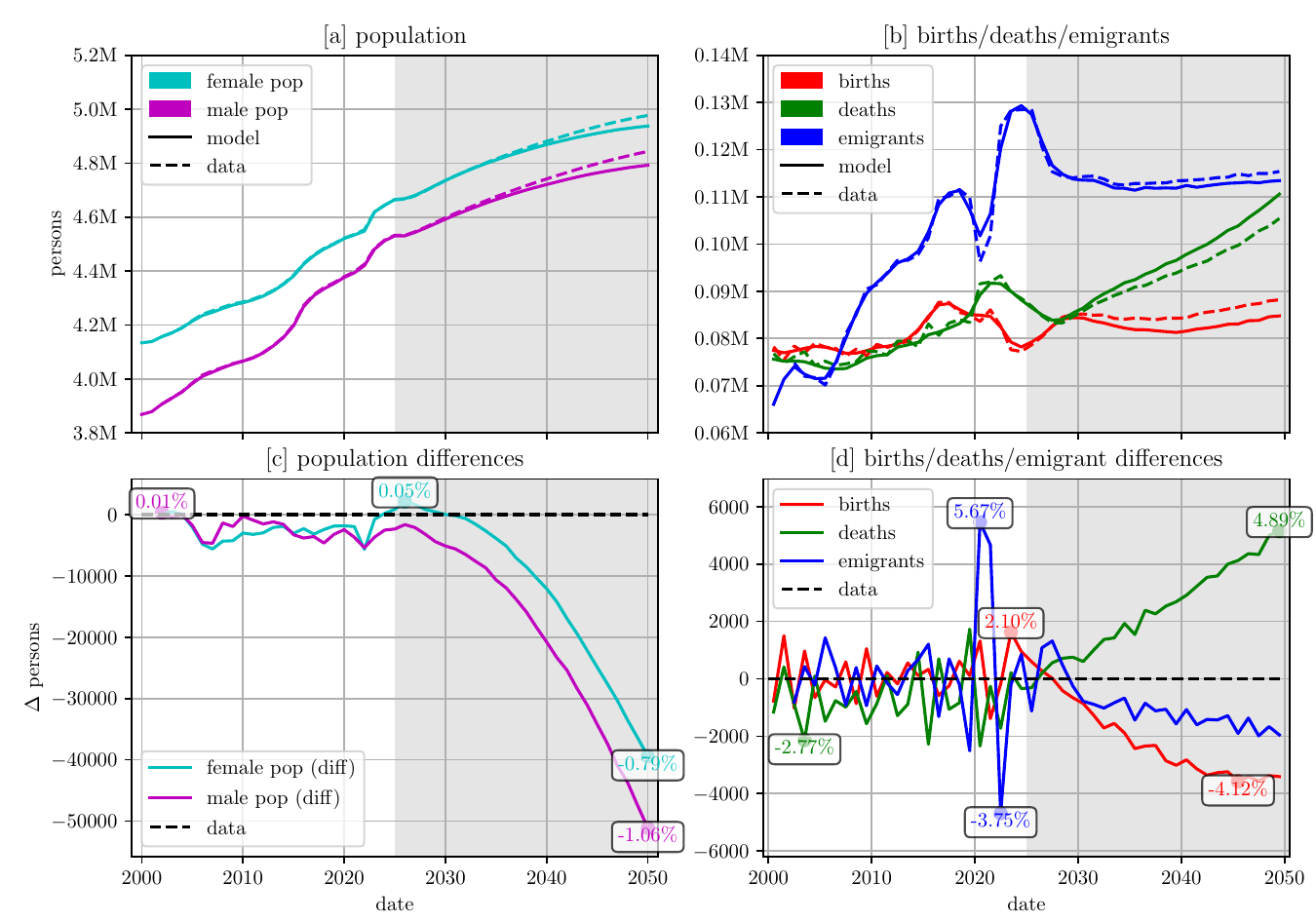}
    \caption{Simulated census (full lines) compared to the reference data (dashed) for different demographic indicators between 2000 and 2050. Panels [a] and [c] show the comparison of the population counts, differentiated by male and female, [b] and [d] show total births, deaths and emigrants. While panels [a] and [b] display total numbers, panels [c] and [d] show the differences together with the maximum and minimum deviations $e_{max}, e_{min}$. The grey background indicates years in which the reference series consists of official forecasts; the white background indicates years with census counts.}
    \label{fig:timeseries}
\end{figure}

In a nutshell, Figure \ref{fig:timeseries} shows the similarity of the simulated census to the actual census, or the census forecast, respectively. The period between 2000 and 2025 stands out with a particularly good fit. During this period, high-resolution census data were used to fit the model. The result curves for the total population are nearly indistinguishable from the reference and within $\pm 0.2\%$, and the births, deaths and emigrants are within $\pm 3\%$ error from the reference. Emigrants show a slightly larger deviation, however, only due to heavy annual fluctuations, which are slightly smoothed by the model. From 2025 to 2050, the results start to diverge from the reference forecast: births and emigrants tend to get underestimated, deaths are slightly overestimated. These differences are mainly due to a potential mismatch between the age distributions of population, births, deaths and migrants, caused by the disaggregation process of the low-resolution forecast data. However, they also originate in structural differences between GEPOC ABM model and the forecasting tool of Statistics Austria.

\begin{figure}
    \centering
    \includegraphics[width=1.0\linewidth]{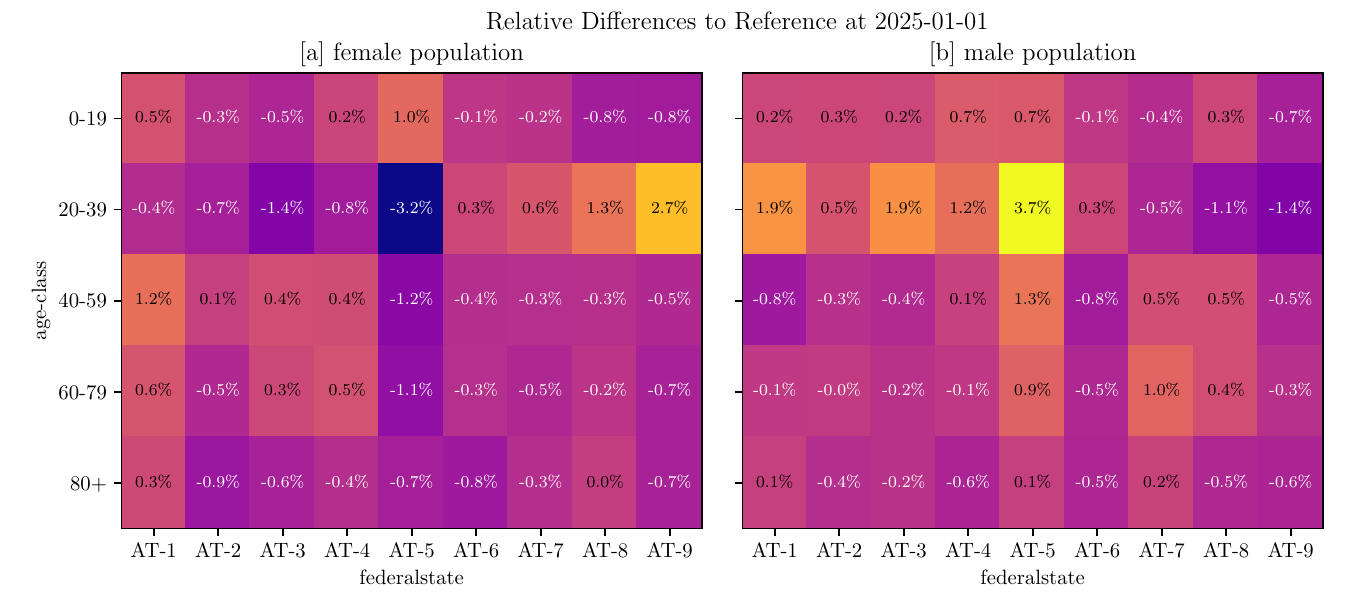}
    \caption{Relative differences between simulated and reference male and female populations on Jan 1\textsuperscript{st} 2025 for different age-classes and the nine federal-states AT-1 to AT-9.}
    \label{fig:heatmap}
\end{figure}

While panels [a] and [c] illustrate a near perfect fit of the overall population until 2025, higher deviations are revealed when looking closer into finer resolutions. Figure \ref{fig:heatmap} shows a detailed picture of the differences between the simulated and actual overall population for Jan 1\textsuperscript{st} 2025, with respect to sex, age-class and federal-state. On this lower level of aggregation, the errors are larger, but with the exception of a few outliers for 20-39-year-olds in Salzburg (AT-5), they remain within $\pm 2\%$ after 25 simulated years. These outliers are due to a slight disruption of the migration in 2022, likely due to the COVID-19 crisis, which could not be fully depicted by the simulation and leads to a temporary unbalance between male and female in this region.

Finally, Table \ref{tbl:deviations} shows the defined error metrics for the overall population on different levels of aggregation. As expected, relative differences tend to become larger with smaller sised cohorts.

\section{Discussion}
In this paper, we have presented several relevant updates to the agent-based population model GEPOC ABM. The concepts presented have contributed to making the model even more valid, flexible and intuitive when applied for a specific population-level case study. Thematically, they range from conceptual modelling, implementation and simulation, to model parametrisation, and we are confident that many of those are well transferable to other models.

The agent's birthday-oriented update scheme leads to the intuitive embedding of demographic processes into any agent-based model, in which these processes are incidental but must nevertheless be taken into account. The corresponding events schedule themselves dynamically and are thus encapsulated and always active in the background. This allows to focus modelling on the processes relevant to the application, e.g. the modelling of contacts and disease dynamics in epidemiological applications.

The synchronisation scheme between agents, which is motivated by co-simulation, offers an intuitive concept for the parallelisation of population-dynamic, time-continuous, agent-based systems. Although enormous progress has been made in recent years in the field of high-performance agent-based simulation (e.g. \cite{breitwieser2021}), solutions for parallelisation are still quite application-specific. In our approach, the problem is already taken into account on the modelling side, which makes potential weaknesses of parallelisation more transparent and explainable. In addition, even without the use of parallel computing, the concept increases the performance of time-continuous ABMs by distributing the effort of sorting event queues among the agents.

The results presented in the previous chapter (Figures \ref{fig:timeseries}, \ref{fig:heatmap}) show how important a sophisticated parameter calculation is for the fit of the model. Tiniest errors in the age-dependent parameters amplify themselves after multiple years of independent simulation. Therefore, the presented level of accuracy can only be achieved with the help of well-thought-out parametrisation methods, e.g. by using the Farr formula for computation of event probabilities instead of a naive division. However, the comparably high deviations for the birth, death, and emigration forecasts also highlight weaknesses of the approaches used for disaggregation of low-resolution data.

Nevertheless, the results show a high degree of consistency with the reference data, making GEPOC ABM a universally applicable tool for answering population-level research questions. The comparatively small number of features offered by the agents –- for example, the basic version of GEPOC ABM does not include households, socio-economic factors or any kind of infrastructure –- does not pose an obstacle here, as (a) not all GEPOC ABM applications actually need these features in the model (modelling paradigm: ``make it as simple as possible, but not simpler''), and (b) it is very easy to add corresponding modules in a modular fashion when needed. In general, the wide range of different application areas, referring to the list stated in the introduction, poses one of the key differences of GEPOC ABM compared to other synthetic population frameworks, which typically focus on one specific application area. For example, the agent-based framework GEMS~\cite{ponge2024standardized} focuses solely on epidemiological applications, the microsimulation model microDems~\cite{fink2020microdems} focuses on economic applications.

In recent years, GEPOC ABM has proven itself not only in a wide variety of applications, but also in different application modes. Classic \underline{agent-based simulation}, where complexity originates from interaction, e.g., as in \cite{bicher_evaluation_2021}, where GEPOC ABM was used as an agent-based epidemic model, is only one of these modes. Sometimes, the model is applied in a static version. Hereby, the agent population is sampled and immediately exported for further processing. This leaves space for \underline{cross-sectional analysis}, where the synthetic population is put in contrast with other data, such as in \cite{schramm_earth_2025}, where it is compared with flooding data. Alternatively, the snapshot can also be used as \underline{input for other models}, such as in \cite{bicher_model-based_2025}, where a GEPOC export was used to generate a spatial-network for a dynamic differential-equation based wastewater model. Finally, sometimes the model is applied dynamically, however, without interaction between the agents -- i.e. the model is used as a \underline{microsimulation model}. Examples are \cite{zauner_microscopic_2019,miksch_evaluation_2017}, where agents were enhanced by certain health-specific properties and dynamically updated without direct interaction. With these examples, we want to emphasise that applications of GEPOC ABM beyond the traditional use of agent-based simulation are not only possible but can also be highly relevant for decision support.

The presented freedom of use of the tool does not only have advantages. Since the model does not have a clearly definable modelling purpose and model boundaries, each application must first be carefully checked for feasibility with regard to the limitations of the underlying population model. On the one hand, the model inherits the limitations of the parameterisation data, which does not only refer to quality, but also to resolution. For example, due to the low resolution of the corresponding datasets, performing simulations far into the future using the open-data parametrisation is only reasonable if slight underestimation of the population compared to the official forecast is acceptable. On the other hand, also the modelling assumptions lead to limitations. In particular, the interaction strategy between the agents makes it important to use an adequate macro time-step length, which is reasonable for the specific application. If chosen too large, the time-lag for interaction and creation of new agents will be too large, rendering the results inaccurate. If chosen too small, computation time will suffer from various unnecessary agent-calls. In general, the comparably long computation can be considered as a particularly important limitation for GEPOC ABM. Depending on the complexity of the application, simulation times are more likely to be in the range of minutes or hours than seconds or microseconds. This drastically limits capabilities for being used within systematic optimisation or calibration loops. Finally, the mentioned static applications of the tool have different requirements for the validity, and therefore come with different limitations. While, for example, sampling of valid location coordinates is hardly important for the dynamic simulation, it is crucial for spatial analysis of a static population snapshot. Currently, we apply a strategy which combines uniform sampling from open census data within geometries (municipality regions) with open raster-data (settlement map). While the concept provides a visibly accurate picture of the population distribution within Austria, one finds weaknesses when zooming in.

We are currently working on reducing these limitations to make the areas of application of GEPOC ABM even broader. This includes, for example, improving assumptions for the disaggregation of low-resolution forecasts, optimising the simulation runtime, or the development of an alternative sampling method based on building-level data instead of a settlement raster-map.

In this paper we have presented various methods that have been primarily responsible for making GEPOC ABM more flexible, valid and efficient in recent years, and thus increasingly used in decision support. These methods are general and can be transferred to any agent-based model with appropriate adjustments.

\section*{Funding}
This research was funded in part by the Austrian Science Fund (FWF) [Grant ID: I 5908-G], the Vienna Science and Technology Fund (WWTF) and the State of Lower Austria [Grant ID: 1047379/LS22071], and the Austrian Research Promotion Agency (FFG), Digitaler Zwilling 2024 [FO999918405].

\bibliographystyle{elsarticle-num} 
\bibliography{references}

\appendix
\section{Appendix}
Table \ref{tbl:deviations} shows the differences between simulation results and source data on different levels of aggregation.
\begin{table}
    \begin{center}
    \begin{scriptsize}
        \begin{tabular}{ccc|c|c}
        region & sex & age & $e_{min} (\text{90\%CI})$ & $e_{max} (\text{90\%CI})$ \\
        \hline
-&-&-&\cellcolor{blue!24}$-0.92\%$ ($-1.0$,$-0.9$)&\cellcolor{red!1}$0.01\%$ ($-0.0$,$0.0$)\\
\hline
-&male&-&\cellcolor{blue!27}$-1.06\%$ ($-1.1$,$-1.0$)&\cellcolor{red!1}$0.01\%$ ($-0.0$,$0.0$)\\
-&female&-&\cellcolor{blue!20}$-0.79\%$ ($-0.9$,$-0.7$)&\cellcolor{red!2}$0.05\%$ ($0.0$,$0.1$)\\
\hline
-&-&$[0,19]$&\cellcolor{blue!47}$-1.85\%$ ($-2.0$,$-1.8$)&\cellcolor{red!24}$0.93\%$ ($0.9$,$1.0$)\\
-&-&$[20,39]$&\cellcolor{blue!53}$-2.09\%$ ($-2.1$,$-2.0$)&\cellcolor{red!9}$0.34\%$ ($0.3$,$0.4$)\\
-&-&$[40,59]$&\cellcolor{blue!66}$-2.63\%$ ($-2.7$,$-2.6$)&\cellcolor{blue!4}$-0.13\%$ ($-0.1$,$-0.1$)\\
-&-&$[60,79]$&\cellcolor{blue!12}$-0.47\%$ ($-0.5$,$-0.4$)&\cellcolor{red!45}$1.77\%$ ($1.7$,$1.8$)\\
-&-&$80^+$&\cellcolor{blue!42}$-1.67\%$ ($-1.8$,$-1.6$)&\cellcolor{red!17}$0.67\%$ ($0.6$,$0.8$)\\
\hline
AT-1&-&-&\cellcolor{blue!19}$-0.74\%$ ($-0.9$,$-0.6$)&\cellcolor{red!9}$0.34\%$ ($0.1$,$0.5$)\\
AT-2&-&-&\cellcolor{blue!42}$-1.67\%$ ($-1.8$,$-1.5$)&\cellcolor{blue!1}$-0.01\%$ ($-0.0$,$0.0$)\\
AT-3&-&-&\cellcolor{blue!16}$-0.61\%$ ($-0.7$,$-0.5$)&\cellcolor{red!5}$0.20\%$ ($0.1$,$0.2$)\\
AT-4&-&-&\cellcolor{blue!38}$-1.49\%$ ($-1.6$,$-1.4$)&\cellcolor{red!7}$0.25\%$ ($0.2$,$0.3$)\\
AT-5&-&-&\cellcolor{blue!1}$-0.01\%$ ($-0.2$,$0.1$)&\cellcolor{red!12}$0.44\%$ ($0.3$,$0.6$)\\
AT-6&-&-&\cellcolor{blue!44}$-1.72\%$ ($-1.8$,$-1.6$)&\cellcolor{blue!7}$-0.25\%$ ($-0.4$,$-0.1$)\\
AT-7&-&-&\cellcolor{blue!44}$-1.73\%$ ($-1.9$,$-1.6$)&\cellcolor{red!10}$0.37\%$ ($0.3$,$0.4$)\\
AT-8&-&-&\cellcolor{blue!52}$-2.08\%$ ($-2.3$,$-1.9$)&\cellcolor{red!5}$0.19\%$ ($0.1$,$0.2$)\\
AT-9&-&-&\cellcolor{blue!12}$-0.47\%$ ($-0.5$,$-0.4$)&\cellcolor{red!9}$0.33\%$ ($0.3$,$0.4$)\\
        \hline
AT-1&-&$[0,19]$&\cellcolor{blue!33}$-2.59\%$ ($-2.9$,$-2.1$)&\cellcolor{red!15}$1.19\%$ ($0.9$,$1.6$)\\
AT-1&-&$[20,39]$&\cellcolor{blue!49}$-3.87\%$ ($-4.3$,$-3.4$)&\cellcolor{red!13}$0.97\%$ ($0.6$,$1.4$)\\
AT-1&-&$[40,59]$&\cellcolor{blue!56}$-4.42\%$ ($-4.7$,$-4.0$)&\cellcolor{red!12}$0.89\%$ ($0.5$,$1.3$)\\
AT-1&-&$[60,79]$&\cellcolor{blue!6}$-0.45\%$ ($-0.6$,$-0.2$)&\cellcolor{red!58}$4.58\%$ ($4.2$,$5.0$)\\
AT-1&-&$80^+$&\cellcolor{blue!23}$-1.80\%$ ($-2.6$,$-1.4$)&\cellcolor{red!24}$1.91\%$ ($1.4$,$2.5$)\\
\hline
AT-2&-&$[0,19]$&\cellcolor{blue!30}$-2.38\%$ ($-2.7$,$-1.9$)&\cellcolor{red!10}$0.78\%$ ($0.6$,$1.0$)\\
AT-2&-&$[20,39]$&\cellcolor{blue!10}$-0.77\%$ ($-1.1$,$-0.4$)&\cellcolor{red!6}$0.42\%$ ($0.1$,$0.8$)\\
AT-2&-&$[40,59]$&\cellcolor{blue!34}$-2.69\%$ ($-2.9$,$-2.4$)&\cellcolor{blue!1}$-0.01\%$ ($-0.1$,$0.1$)\\
AT-2&-&$[60,79]$&\cellcolor{blue!17}$-1.31\%$ ($-1.6$,$-1.0$)&\cellcolor{blue!3}$-0.20\%$ ($-0.3$,$0.1$)\\
AT-2&-&$80^+$&\cellcolor{blue!39}$-3.06\%$ ($-3.5$,$-2.5$)&\cellcolor{blue!1}$-0.01\%$ ($-0.5$,$0.4$)\\
\hline
AT-3&-&$[0,19]$&\cellcolor{blue!41}$-3.24\%$ ($-3.5$,$-3.1$)&\cellcolor{red!11}$0.83\%$ ($0.7$,$1.0$)\\
AT-3&-&$[20,39]$&\cellcolor{blue!44}$-3.44\%$ ($-3.6$,$-3.3$)&\cellcolor{red!7}$0.49\%$ ($0.3$,$0.7$)\\
AT-3&-&$[40,59]$&\cellcolor{blue!33}$-2.57\%$ ($-2.7$,$-2.4$)&\cellcolor{red!4}$0.27\%$ ($0.1$,$0.4$)\\
AT-3&-&$[60,79]$&\cellcolor{blue!6}$-0.42\%$ ($-0.5$,$-0.4$)&\cellcolor{red!55}$4.36\%$ ($4.2$,$4.5$)\\
AT-3&-&$80^+$&\cellcolor{blue!24}$-1.84\%$ ($-2.2$,$-1.5$)&\cellcolor{red!22}$1.69\%$ ($1.6$,$2.0$)\\
\hline
AT-4&-&$[0,19]$&\cellcolor{blue!46}$-3.63\%$ ($-4.0$,$-3.3$)&\cellcolor{red!11}$0.83\%$ ($0.8$,$0.9$)\\
AT-4&-&$[20,39]$&\cellcolor{blue!41}$-3.20\%$ ($-3.4$,$-3.1$)&\cellcolor{red!8}$0.63\%$ ($0.6$,$0.7$)\\
AT-4&-&$[40,59]$&\cellcolor{blue!36}$-2.85\%$ ($-3.0$,$-2.7$)&\cellcolor{red!8}$0.63\%$ ($0.5$,$0.8$)\\
AT-4&-&$[60,79]$&\cellcolor{blue!8}$-0.60\%$ ($-0.7$,$-0.5$)&\cellcolor{red!30}$2.40\%$ ($2.2$,$2.6$)\\
AT-4&-&$80^+$&\cellcolor{blue!31}$-2.45\%$ ($-2.8$,$-2.1$)&\cellcolor{red!1}$0.06\%$ ($-0.2$,$0.4$)\\
\hline
AT-5&-&$[0,19]$&\cellcolor{red!2}$0.14\%$ ($-0.2$,$0.5$)&\cellcolor{red!34}$2.66\%$ ($2.0$,$3.3$)\\
AT-5&-&$[20,39]$&\cellcolor{blue!40}$-3.16\%$ ($-3.5$,$-2.8$)&\cellcolor{red!6}$0.43\%$ ($0.2$,$0.7$)\\
AT-5&-&$[40,59]$&\cellcolor{blue!14}$-1.11\%$ ($-1.5$,$-0.9$)&\cellcolor{red!9}$0.66\%$ ($0.5$,$0.8$)\\
AT-5&-&$[60,79]$&\cellcolor{blue!8}$-0.63\%$ ($-0.7$,$-0.5$)&\cellcolor{red!29}$2.24\%$ ($2.1$,$2.4$)\\
AT-5&-&$80^+$&\cellcolor{blue!28}$-2.23\%$ ($-2.7$,$-1.6$)&\cellcolor{red!21}$1.64\%$ ($1.2$,$2.2$)\\
\hline
AT-6&-&$[0,19]$&\cellcolor{blue!37}$-2.89\%$ ($-3.1$,$-2.6$)&\cellcolor{red!9}$0.71\%$ ($0.5$,$1.0$)\\
AT-6&-&$[20,39]$&\cellcolor{blue!46}$-3.61\%$ ($-3.8$,$-3.5$)&\cellcolor{red!5}$0.35\%$ ($0.2$,$0.6$)\\
AT-6&-&$[40,59]$&\cellcolor{blue!46}$-3.61\%$ ($-3.8$,$-3.4$)&\cellcolor{blue!5}$-0.39\%$ ($-0.5$,$-0.3$)\\
AT-6&-&$[60,79]$&\cellcolor{blue!8}$-0.60\%$ ($-0.8$,$-0.5$)&\cellcolor{red!19}$1.47\%$ ($1.2$,$1.6$)\\
AT-6&-&$80^+$&\cellcolor{blue!34}$-2.70\%$ ($-3.0$,$-2.4$)&\cellcolor{red!5}$0.40\%$ ($0.1$,$0.8$)\\
\hline
AT-7&-&$[0,19]$&\cellcolor{blue!61}$-4.86\%$ ($-5.2$,$-4.5$)&\cellcolor{red!15}$1.14\%$ ($1.0$,$1.2$)\\
AT-7&-&$[20,39]$&\cellcolor{blue!38}$-2.96\%$ ($-3.3$,$-2.7$)&\cellcolor{red!4}$0.32\%$ ($0.2$,$0.5$)\\
AT-7&-&$[40,59]$&\cellcolor{blue!33}$-2.62\%$ ($-2.9$,$-2.3$)&\cellcolor{red!4}$0.30\%$ ($0.2$,$0.3$)\\
AT-7&-&$[60,79]$&\cellcolor{blue!5}$-0.33\%$ ($-0.5$,$-0.1$)&\cellcolor{red!17}$1.29\%$ ($1.1$,$1.5$)\\
AT-7&-&$80^+$&\cellcolor{blue!18}$-1.40\%$ ($-1.9$,$-0.6$)&\cellcolor{red!17}$1.28\%$ ($1.1$,$1.4$)\\
\hline
AT-8&-&$[0,19]$&\cellcolor{blue!49}$-3.88\%$ ($-4.4$,$-3.5$)&\cellcolor{red!8}$0.60\%$ ($0.5$,$0.8$)\\
AT-8&-&$[20,39]$&\cellcolor{blue!23}$-1.78\%$ ($-2.2$,$-1.5$)&\cellcolor{red!5}$0.38\%$ ($0.2$,$0.7$)\\
AT-8&-&$[40,59]$&\cellcolor{blue!52}$-4.16\%$ ($-4.5$,$-3.8$)&\cellcolor{red!2}$0.16\%$ ($0.1$,$0.4$)\\
AT-8&-&$[60,79]$&\cellcolor{blue!5}$-0.37\%$ ($-0.6$,$-0.1$)&\cellcolor{red!4}$0.29\%$ ($0.1$,$0.6$)\\
AT-8&-&$80^+$&\cellcolor{blue!33}$-2.58\%$ ($-3.0$,$-2.3$)&\cellcolor{red!11}$0.81\%$ ($0.2$,$1.6$)\\
\hline
AT-9&-&$[0,19]$&\cellcolor{blue!10}$-0.73\%$ ($-1.0$,$-0.5$)&\cellcolor{red!41}$3.21\%$ ($2.9$,$3.4$)\\
AT-9&-&$[20,39]$&\cellcolor{blue!5}$-0.36\%$ ($-0.4$,$-0.3$)&\cellcolor{red!18}$1.37\%$ ($1.3$,$1.5$)\\
AT-9&-&$[40,59]$&\cellcolor{blue!24}$-1.91\%$ ($-2.1$,$-1.8$)&\cellcolor{blue!6}$-0.45\%$ ($-0.6$,$-0.3$)\\
AT-9&-&$[60,79]$&\cellcolor{blue!9}$-0.72\%$ ($-0.9$,$-0.6$)&\cellcolor{blue!1}$-0.07\%$ ($-0.2$,$0.1$)\\
AT-9&-&$80^+$&\cellcolor{blue!26}$-2.03\%$ ($-2.3$,$-1.6$)&\cellcolor{red!19}$1.45\%$ ($1.2$,$1.8$)\\
        \end{tabular}
\end{scriptsize}
\end{center}
    \caption{Relative minimum and maximum deviations between the simulated census and the reference data for the total population on Jan 1\textsuperscript{st} between 2000 and 2050 on different levels of aggregation. The depth of the colour indicates by how much the values are under- or overestimated.}
    \label{tbl:deviations}
\end{table}





\end{document}